\documentclass[prd,byrevtex
,preprint
,showkeys
,showpacs
,amsfonts,amsmath,amssymb
]{revtex4}
\usepackage[pdftex]{graphicx}
\usepackage{float}
\usepackage{color}
\begin{document}
\title{Large scale quantum entanglement in de Sitter spacetime}
\author{Akira Matsumura}
\email{matsumura.akira@h.mbox.nagoya-u.ac.jp}
\author{Yasusada Nambu}
\email{nambu@gravity.phys.nagoya-u.ac.jp}
\affiliation{Department of Physics, Graduate School of Science, Nagoya 
University, Chikusa, Nagoya 464-8602, Japan}
\date{May 29, 2018} 
%
\begin{abstract}
  We investigate quantum entanglement between two symmetric spatial
  regions in de Sitter space with the Bunch-Davies vacuum.  As a
  discretized model of the scalar field for numerical simulation, we
  consider a harmonic chain model. Using the coarse-grained variables
  for the scalar field, it is shown that the multipartite
    entanglement on the superhorizon scale exists by checking the
    monogamy relation for the negativity which quantifies the
    entanglement between the two regions. Further, we consider the
  continuous limit of this model without coarse-graining and
  find that non-zero values of the logarithmic negativity exist even
  if the distance between two spatial regions is larger than the
  Hubble horizon scale.
\end{abstract}
\maketitle

\tableofcontents
\section{Introduction}
Quantum entanglement is an interesting aspect of quantum physics,
which has recently received remarkable attention from various fields. 
As is well known in the quantum theory or quantum
information theory, quantum entanglement represents a non-local correlation and leads to
the violation of the Bell-CHSH
inequality~\cite{Bell1964b,Clauser1969b}. This quantum correlation is
needed as a resource to carry out some protocols, e.g. quantum
teleportation, superdense coding, quantum error correction and so on
\cite{Nielsen2007}. For quantum many-body systems or quantum field
theories, the entanglement of the ground state or vacuum state
characterizes the structure of its wave function. In a previous work
by S.~Marcovitch \textit{et al.}~\cite{Marcovitch2009a}, the quantum
entanglement of a free Klein-Gordon field in the 1+1-dimensional
Minkowski spacetime was examined in terms of the logarithmic
negativity which is a useful measure to quantify the entanglement for
a mixed state~\cite{Vidal2002a}. They focused on two spatially
separated regions and numerically investigated the logarithmic
negativity for the Minkowski vacuum as a function of the ratio of the
separation between the regions to the size of the each region. It was
shown that the logarithmic negativity decays exponentially as the
ratio becomes large. This behavior is consistent with the
Reeh-Schlieder theorem \cite{Reeh1961} which implies that the quantum
entanglement persists for all scales for the Minkowski vacuum.

The nonlocal quantum correlation also was examined for
cosmological
situations~\cite{Maldacena2015,Kanno2015,Kanno2014,Nambu2008,Nambu2011}.
In a previous work by Y.~Nambu~\cite{Nambu2008}, the quantum
entanglement of a massless scalar field in de Sitter space was
investigated using the spatially coarse grained scalar field. It was
revealed that the logarithmic negativity for the coarse-grained field
vanishes, when the physical size of each region exceeds the Hubble
horizon and becomes causally disconnected due to the de Sitter
expansion. This behavior of entanglement is consistent with the
scenario of quantum to classical transition of the primordial
fluctuation generated by inflationary expansion in the early universe
\cite{Guth1985, Polarski1996, Lesgourgues1997,Kiefer1998, Kiefer2007,
  Burgess2008}. The similar transition to zero negativity state is
known for a model of 1+1-dimensional harmonic chain with a finite
temperature \cite{Anders2008}. For sufficiently high
  temperatures, the negativity between two spatial regions becomes
zero. Qualitatively, this transition can be understood as follows;
above the critical temperature, the wavelength of the thermal
fluctuations becomes shorter than the lattice spacing and the quantum
correlation between adjacent lattice sites is destroyed by the thermal
fluctuations. Concering the quantum field in de Sitter space, the
effective comoving lattice spacing for the coarse-grained field
becomes larger than the wavelength of quantum fluctuations which is
equal to the Hubble length of de Sitter space, and the negativity
becomes zero.  We expect that appearance of zero negativity state is
related to the coarse-graining treatment of the quantum field. As
mentioned above, for the Minkowski vacuum in quantum field theory, the
Reeh-Schrieder theorem is known and the theorem also holds for
thermal states \cite{Jaekel2000, Narnhofer2005}. It seems that there
is a difference between the feature of entanglement in the
coarse-grained field and its continuous limit. Thus for complete
understanding the property of the entanglement in de Sitter space, we need to
analyze the quantum entanglement between two spatial regions using
both a model with coarse-graining and one without coarse-graining.

In this paper, we use a 1+1-dimensional lattice model of a massless
scalar field whose mode equation is equivalent to that of the
1+3-dimensional de Sitter spacetime and numerically evaluate the
entanglement between two spatial regions.  As a quantum state, we
assume the Bunch-Davies vacuum, which is a vacuum state in de Sitter
spacetime and corresponds to the Minkowski vacuum in the far remote
past. We introduce coarse-grained variables in the lattice model and
the negativity between two spatial regions is calculated.  To
understand the connection between the coarse-grained system and the
continuous one, the effect of multipartite entanglement is considered
using the monogamy relation for the negativity. We also compute the
logarithmic negativity between the two regions without the
coarse-graining and its behavior is discussed especially focusing on
the continuous limit and the existence of the Hubble horizon.

This paper is organized as follows: we introduce our 1+1-dimensional
lattice model of a minimal coupled massless scalar field in de Sitter
spacetime and the entanglement measures (the negativity and the
logarithmic negativity) for the Gaussian state in Sec.~II.  In
Sec.~III, we define the coarse-grained variables in the lattice system
and calculate the negativity numerically. Then we discuss the
monogamous property of the negativity. Also, we present our main
numerical result of the logarithmic negativity without coarse-graining
and provide the fitting formula for the super horizon scale. Sec.~IV
is devoted to summary and conclusion.

\section{Harmonic chain model and entanglement measures for  Gaussian
  state}

The Hamiltonian for a minimal coupled massless scalar field $q$ in de
Sitter spacetime with a spatially flat slice is given by
\begin{equation}
\mathcal{H}=\int d^{3} x  \frac{1}{2} \left[p^{2}+(\partial_{i}
q)^{2}+\frac{1}{a}\frac{d a}{d\eta}\,(qp+pq) \right], \quad
a=-\frac{1}{H\eta}, 
\end{equation}
where $a$ is the scale factor,  $\eta<0$
is the conformal time and  $H$ represents the Hubble constant. For
simplicity of numerical analysis, we assume that the field depends
only on the conformal time $\eta$ and one spatial coordinate. By
introducing a lattice spacing $\Delta x$ of the spatial direction, the
dimensionless form of the Hamiltonian of this model (harmonic
chain)~\cite{Nambu2008} is expressed as
\begin{equation}
\mathcal{H}=\sum_{j=1}^{N} \left[ \frac{1}{2}\, p_{j}^{2}+q_{j}^{2}-\alpha\,
 q_{j} q_{j-1}+\frac{1}{2a}\frac{da}{d\tau}\,(p_{j}q_{j}+q_{j}p_{j}) \right], 
\end{equation}
where we impose a periodic boundary condition on $q_{j}$ and $p_{j}$
to respect the translational invariance of the model. $N$ denotes the
total number of lattice sites. $\tau$ and $\alpha$ are the
dimensionless conformal time and the IR cutoff parameter,
respectively. These two parameters are given by
\begin{equation}
\tau=\eta/\Delta x, \quad (m \Delta x)^{2}=2(1-\alpha),
\end{equation}
where  $m$ is the mass of the
scalar field corresponding to the IR cutoff. We quantize this model as follows:
\begin{equation}
\hat{q}_{j}=\frac{1}{\sqrt{N}} \sum_{k=0}^{N-1} \left[ \hat{a}_{k}\,
  f_{k}+\hat{a}_{N-k}^{\dagger}\, f_{k}^{*} \right]e^{i\theta_{k}j}, \quad
\hat{p}_{j}=\frac{-i}{\sqrt{N}} \sum_{k=0}^{N-1} 
\left[ \hat{a}_{k}\, g_{k}-\hat{a}_{N-k}^{\dagger}\, g_{k}^{*} \right]e^{i\theta_{k}j},
\end{equation}
where $\theta_{k}=2\pi k/N$. $\hat{a}_{k}$ and $\hat{a}_{k}^{\dagger}$
are annihilation and creation operators, which obey the
commutation relations
\begin{equation}
\left[\hat{a}_{k},\hat{a}_{k'}^{\dagger} \right]=\delta_{kk'}, \quad
\left[\hat{a}_{k},\hat{a}_{k'}
\right]=\left[\hat{a}_{k}^{\dagger},\hat{a}_{k'}^{\dagger}
 \right]=0.
\end{equation}
The mode functions $f_{k}$ and $g_{k}$ satisfy
\begin{equation}
\ddot{f}_{k}+\left(\omega^{2}_{k}-\frac{\ddot{a}}{a} \right)f_{k}=0,\quad
f_{k}\dot{f}^{*}_{k}-\dot{f}_{k}f^{*}_{k}=i,\quad
g_{k}=i \left(\dot{f}_{k}-\frac{\dot{a}}{a} f_{k} \right),
\end{equation}
where ``$\cdot$" is the derivative with respect to $\tau$ and
$\omega_k^2=2(1-\alpha\cos\theta_k)$. Although this lattice model is
introduced in the 1+1-dimensional spacetime, we assume that the equation
of the mode functions has the same form as that in the 1+3-dimensional de Sitter space. 

For the investigation of quantum entanglement of this system with the
Bunch-Davies vacuum which belongs to a family of Gaussian
states, we present a brief review of the negativity and the
logarithmic negativity for a Gaussian state. Let us consider a phase
space composed of canonical variables
$\{\hat{q}_{j},\hat{p}_{j} \}_{j=1,...,N}$. The canonical commutation
relations are
\begin{equation}
\left[ \hat{R}_{j}, \hat{R}_{k} \right]=i \Omega_{jk}, \quad
\Omega=\bigoplus_{j=1}^{N} J,\quad
J=\begin{bmatrix}
      0 & 1  \\
    -1 & 0 
   \end{bmatrix},
\end{equation}
where $\hat{R}_{j}$ represent canonical variables defined by
\begin{equation}
\hat{R}=\left[ \hat{q}_{1}, \hat{p}_{1},...,\hat{q}_{N},\hat{p}_{N} \right]^{T}.
\end{equation}
A Gaussian state $\hat\rho$ is determined by its first  moment 
$\langle\hat R_j\rangle=\mathrm{Tr}[\hat{\rho}\, \hat{R}_{j}]$ and the
covariance matrix 
\begin{equation}
V_{jk}=\frac{1}{2}\,\mathrm{Tr}\left[ \left(\Delta \hat{R}_{j}\, \Delta 
\hat{R}_{k}+
\Delta \hat{R}_{k}\, \Delta \hat{R}_{j} \right)\hat{\rho} \right],\quad
\Delta \hat{R}_{j}=\hat{R}_{j}-\langle \hat R_{j}\rangle.
\end{equation}
The negativity $\mathcal{N}$ of a bipartite Gaussian state
$\hat{\rho}_{AB}$ is given by using the symplectic eigenvalues of the
partially transposed covariance matrix $\tilde{V}_{AB}$
\cite{Simon2000} obtained from $V_{AB}$ by replacing $\hat{p}_{jA}$
with $-\hat{p}_{jA}$ \cite{Marcovitch2009a}:
\begin{equation}
{\mathcal{N}}=\frac{1}{2}\Bigl(\prod_{j=1}^{N} \frac{1}{\min \left(2\tilde{\nu}_{j},1\right)}-1 \Bigr),
\end{equation}
where ${\tilde \nu}_{j}$ are positive eigenvalues of
$i\, \Omega \,\tilde{V}_{AB}$. The sufficient condition for the
entangled state is that the negativity $\mathcal{N}$ of the state does
not vanish. The logarithmic negativity $E_{\mathcal{N}}$ is defined by
the negativity as 
\begin{equation}
E_{\mathcal{N}}=\ln \left[ 2{\mathcal{N}}+1 \right].
\end{equation}
The logarithmic negativity $E_{\mathcal{N}}$ provides an upper bound
of the distillable entanglement (the number of the Bell pairs
extractable from a bipartite state) \cite{Vidal2002a,Plenio2005}, and if
$E_{\mathcal{N}}$ is nonzero then the bipartite state is
entangled. However, there exists an entangled state whose the
entanglement of distillation vanishes and such a state is called a
bound entangled state. Fortunately, no bound entangled state exists
for a bipartite Gaussian state with $E_{\mathcal{N}}=0$
\cite{Giedke2001}.

To compute the negativity and the logarithmic negativity, we need the two-point functions of canonical variables on each site for the vacuum state. The correlation functions of the vacuum state are 
\begin{align}
&\frac{1}{2} \langle 0|(\hat{q}_{j}\hat{q}_{l}+ \hat{q}_{j}\hat{q}_{l})|0\rangle =
\frac{1}{N} \sum_{k=0}^{N-1} |f_{k}|^{2} \cos \left[ \theta_{k}(j-l)
  \right],   
\\
&\frac{1}{2} \langle 0|(\hat{p}_{j}\hat{p}_{l}+\hat{p}_{l}\hat{p}_{j})|0\rangle=
  \frac{1}{N} \sum_{k=0}^{N-1}|g_{k}|^{2} \cos \left[ \theta_{k}(j-l)
  \right], 
\\ 
&\frac{1}{2} \langle 0|(\hat{q}_{j}\hat{p}_{l}+\hat{p}_{l}\hat{q}_{j})|0\rangle=
  \frac{1}{N}
  \sum_{k=0}^{N-1}\frac{i}{2}(f_{k}g^{*}_{k}-f^{*}_{k}g_{k} )\cos
  \left[ \theta_{k}(j-l) \right].
\label{3-1}
\end{align}
We choose the mode functions $f_{k}$ and $g_{k}$ which correspond to the Bunch-Davies vacuum as follows: 
\begin{equation}
f_{k}=\frac{1}{\sqrt{2\omega_{k}}} \left(1+\frac{1}{i\omega_{k}\tau}
\right)e^{-i\omega_{k} \tau},\quad
g_{k}=\sqrt{\frac{\omega_{k}}{2}}\, e^{-i\omega_{k} \tau}.
\end{equation}
 In the continuous limit of the lattice model, the correlation functions of the Bunch-Davies vacuum are given as
\begin{align}
\frac{1}{2} \langle 0|(\hat{q}(x,\eta)\hat{q}(y,\eta)+ \hat{q}(y,\eta)\hat{q}(x,\eta))|0\rangle 
&= \int^{\infty}_{-\infty} \frac{dk}{2\pi} |f_{k}|^{2} \cos [k(x-y)] \nonumber \\
&\sim \frac{1}{2\pi}(-\gamma-\ln[m|x-y|])+\frac{1}{2\pi (m\eta)^{2}}, \label{16}
\\
\frac{1}{2} \langle 0|(\hat{p}(x,\eta)\hat{p}(y,\eta)+\hat{p}(y,\eta)\hat{p}(x,\eta))|0\rangle
&=\int^{\infty}_{-\infty} \frac{dk}{2\pi} |g_{k}|^{2} \cos [k(x-y)] \nonumber \\
&\sim -\frac{1}{2\pi |x-y|^{2}}, \label{17}
\\ 
\frac{1}{2} \langle 0|(\hat{q}(x,\eta)\hat{p}(y,\eta)+\hat{p}(y,\eta)\hat{q}(x,\eta))|0\rangle
&=
 \int^{\infty}_{-\infty} \frac{dk}{2\pi}\frac{i}{2}(f_{k}g^{*}_{k}-f^{*}_{k}g_{k} )\cos [k(x-y)] \nonumber \\
&\sim \frac{1}{2\pi \eta}(-\gamma-\ln [m|x-y|]), \label{18}
\end{align} 
where the above approximated formulas are obtained for $m|x-y| \ll 1$,
and $\gamma$ is the Euler constant. The two-point correlation of our effective
model in a 1+1-dimensional spacetime decrease with the distance $|x-y|$
more slowly compared to that in the 1+3-dimensional de Sitter space
case.

\section{Behavior of entanglement}
Because the scalar field is a many-body system, 
we expect that the multipartite entanglement is a key property to
understand behavior of entanglement between two spatial regions in the
de Sitter space. For this purpose, we introduce the coarse-grained
field and the monogamy inequality of entanglement to quantify the
multipartite entanglement of the scalar field.

\subsection{Coarse-grained field and entanglement monogamy}
\begin{figure}[H]
\centering
    \includegraphics[clip,width=0.8\linewidth]{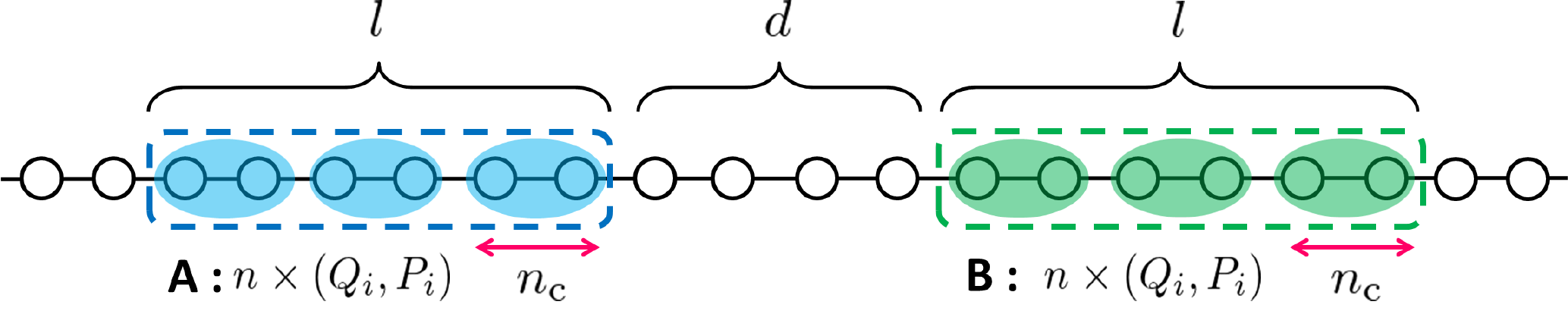}
    \caption{Two spatial symmetric regions A and B in the harmonic
      chain. $l$ is the comoving size of each region and $d$ is the
      comoving distance between the two regions. $n_{\rm{c}}$ is the
      coarse-graining size for each system and $n$ are the number of
      modes of the coarse-grained variables included in each region.}
    \label{fig1}
\end{figure}

To focus on the behavior of the multipartite
entanglement, we introduce the coarse-grained variables
$(\hat{Q}_{i},\hat{P}_{i})$ as follows:
\begin{equation}
\hat{Q}_{i}=\frac{1}{\sqrt{n_{\rm{c}}}} \sum_{j=0}^{n_{\rm{c}}-1} \hat{q}_{n_{\rm{c}}i+j},~\hat{P}_{i}=\frac{1}{\sqrt{n_{\rm{c}}}} \sum_{j=0}^{n_{\rm{c}}-1} \hat{p}_{n_{\rm{c}}i+j},
\label{}
\end{equation} 
where we denote $n_{\rm{c}}$ as the coarse-graining size of the
canonical variables. The coarse-grained variables satisfy the
canonical commutation relations
\begin{equation}
\Bigl[\hat{Q}_{i},~\hat{P}_{j} \Bigr]=i \delta_{ij},~\Bigl[\hat{Q}_{i},~\hat{Q}_{j} \Bigr]=\Bigl[\hat{P}_{i},~\hat{P}_{j} \Bigr]=0.
\label{}
\end{equation} 
We introduce two spatial symmetric regions A and B by choosing the following coarse-grained variables in the harmonic chain: 
\begin{equation}
\hat{R}^{A}=\left[ \hat{Q}_{1}, \hat{P}_{1},...,\hat{Q}_{n},\hat{P}_{n} \right]^{T},~\hat{R}^{B}=\left[ \hat{Q}_{n+1+d/n_{\rm{c}}}, \hat{P}_{n+1+d/n_{\rm{c}}},...,\hat{Q}_{2n+d/n_{\rm{c}}},\hat{P}_{2n+d/n_{\rm{c}}} \right]^{T}
\end{equation}
where $n$ is the number of modes of the coarse-grained variables
contained in each region, and $d$ is the comoving separation between A
and B. The comoving size of each region is denoted as
$l=n\times n_{\rm{c}}$ (FIG.~\ref{fig1}). 
\noindent
Since the vacuum state is
Gaussian, the quantum bipartite entanglement can be completely
characterized by the covariance matrix defined as
\begin{equation}
V_{AB}=\begin{bmatrix}
      A & C  \\
    C^{T} & B 
      \end{bmatrix},\quad
      A^{T}=A, \quad
      B^{T}=B
\end{equation}
where $A, B, C$ are $2n\times 2n$ matrices
given by
\begin{align}
A_{ij}=\frac{1}{2} \langle 0| \bigl(\hat{R}^{A}_{i}\hat{R}^{A}_{j}+\hat{R}^{A}_{j}\hat{R}^{A}_{i}\bigr) |0 \rangle, \nonumber \\ 
B_{ij}=\frac{1}{2} \langle 0| \bigl(\hat{R}^{B}_{i}\hat{R}^{B}_{j}+\hat{R}^{B}_{j}\hat{R}^{B}_{i}\bigr) |0 \rangle, \nonumber \\ 
C_{ij}=\frac{1}{2} \langle 0| \bigl(\hat{R}^{A}_{i}\hat{R}^{B}_{j}+\hat{R}^{B}_{j}\hat{R}^{A}_{i} \bigr)|0 \rangle,
\end{align}
with $\langle 0|\hat{R}_{i}^{(A,B)} |0 \rangle=0$.
In Ref.~\cite{Nambu2008}, the logarithmic negativity between A and B
were calculated with the number of each mode $n=1$, which corresponds to assigning a pair of
canonical variables to each region.

\begin{figure}[H]
 \begin{minipage}{0.5\hsize}
   \centering
   \includegraphics[width=0.95\linewidth]{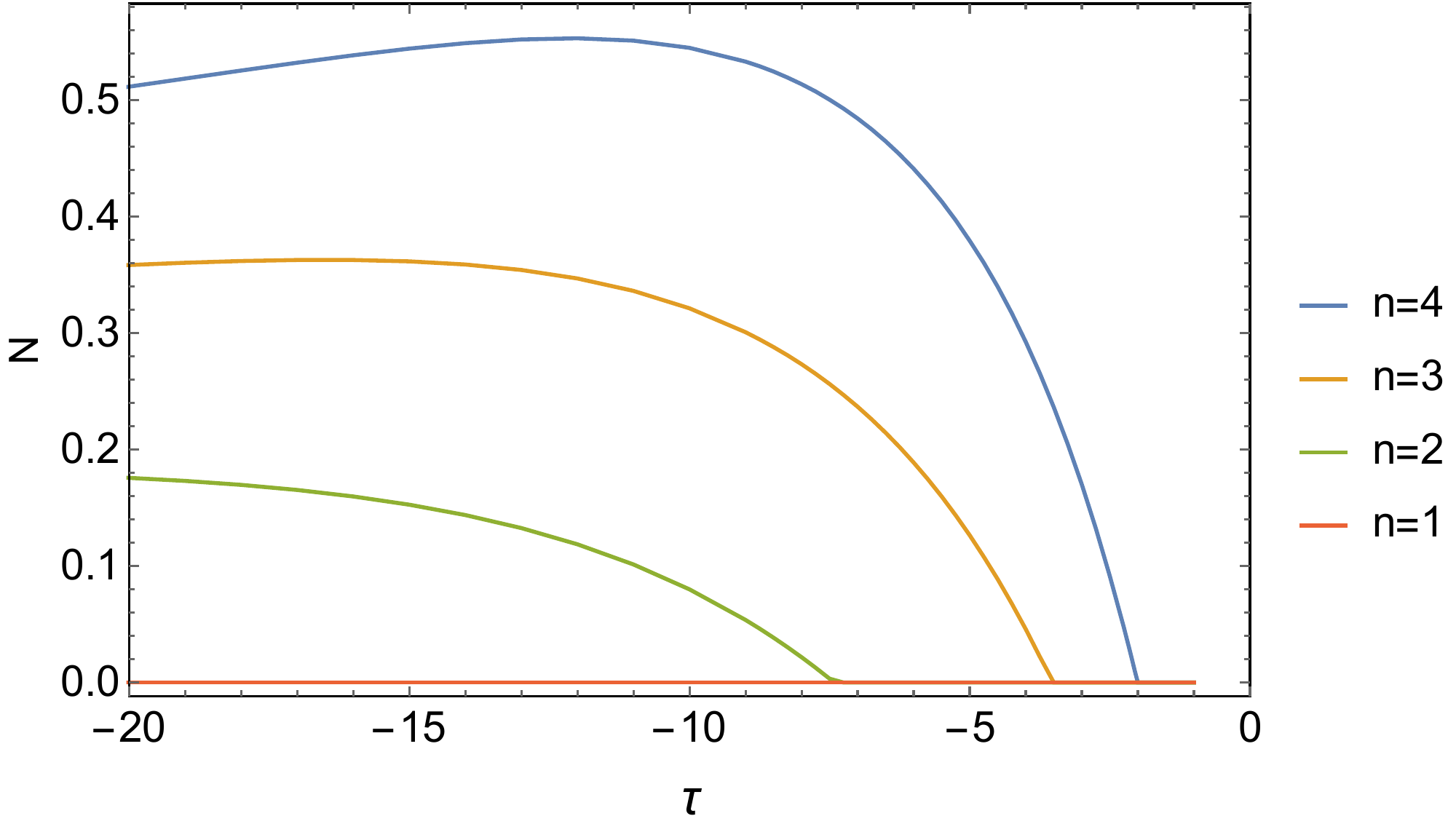}
 \end{minipage}
 \begin{minipage}{0.5\hsize}
   \centering
   \includegraphics[width=0.98\linewidth]{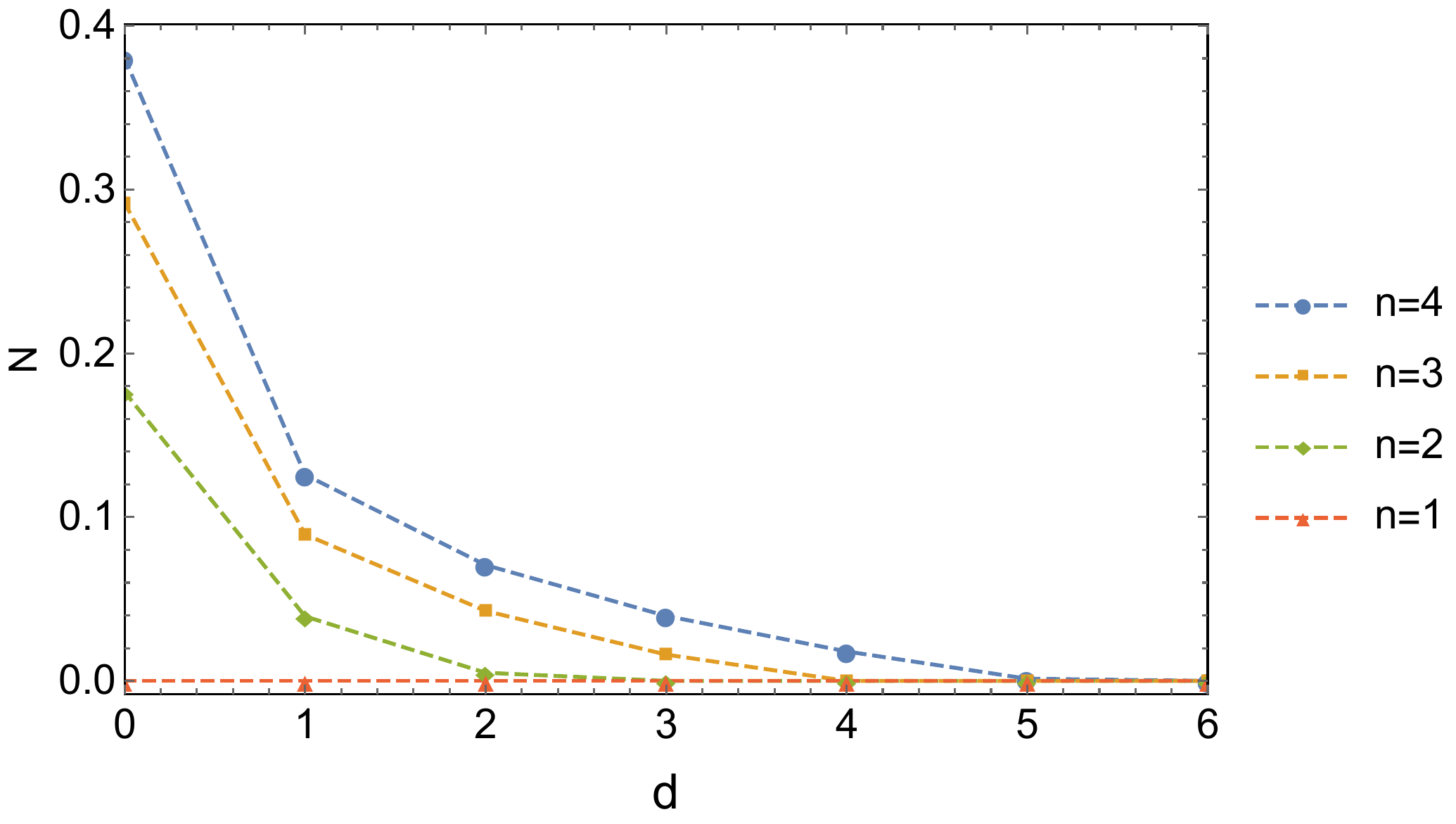}
 \end{minipage}
 \caption{The negativity with different size of the coarse-graining
    . The left panel: the negativity ${\mathcal{N}}$ as a function
   of $\tau$ with $d=0$ and $l=60$. The right panel: the negativity
   ${\mathcal{N}}$ as a function of $d$ for $\tau=-80$.}
    \label{fig2}
\end{figure}

The following results are based on the numerical calculation with the
number of lattice sites $N=2\times 10^{4}$ and the IR cutoff parameter
$\alpha=1-10^{-12}$. FIG. \ref{fig2} shows behavior of the
  negativity with different size of coarse-graining. The left panel
of FIG. \ref{fig2} presents the time evolution of the negativity with
fixed the comoving distance $d=0$ and the comoving size $l=60$. As the
number of modes of the coarse-grained variables $n$ in the system A
increases, the time at which the negativity vanishes becomes
later. The right panel of FIG. \ref{fig2} gives the distance
dependence of the negativity with fixed $l=60$ and $\tau=-80$. For a
larger $n$, the negativity increases and vanishes at a larger distance
$d$.
\noindent 
To
compare the previous work \cite{Nambu2008} with our results, we
introduce
\begin{equation}
l_p=-l/\tau,\quad
d_p=-d/\tau,
\end{equation}
where $l_p$ and $d_p$ represent the proper (physical) size of each
region and the distance between the two regions in the unit of the
Hubble length $H^{-1}$, respectively. In Ref. \cite{Nambu2008}, it is
found that the quantum entanglement between A and B with the
number of each mode $n=1$ disappears when the proper size of each region
is comparable to the Hubble horizon, that is, at $l_{\rm{p}}=1$. This
means that the quantum fluctuation for the super horizon scale behaves
classically in terms of bipartite entanglement. However, according to
the left panel of FIG. \ref{fig2}, we find that the quantum
entanglement with the number of each mode $n\ge2$ in the system A and B does not vanish even if
the proper size of each region $l_{\rm{p}}$ is larger than $1$ (for
example, in the case $n=2$, the negativity is nonzero at
$\tau=-10, l=60$, that is, $l_{\rm{p}}=6$) . Hence we expect there
exists the multipartite entanglement even for the quantum fluctuation
in the super horizon scale. Also, in the right panel of
FIG. \ref{fig2}, the maximum distance that the negativity exists
increases monotonically as $n$ increases (in Ref. \cite{Nambu2008}, the
negativity vanishes trivially for $d\geq1$). It seems that the
multipartite entanglement between casually disconnected regions also
survives for a larger $n$ (in the following section B, we clarify the distance dependence
 for the continuous limit of our model).

To get a clear intuition for the behavior of FIG. \ref{fig2}, we
introduce the monogamy relation of the negativity for this model. We
consider a tripartite state $\rho_{ABC}$ and the negativity
${\mathcal{N}}_{AB|C}$ between AB and C. It is conjectured that
${\mathcal{N}}_{A|C}, {\mathcal{N}}_{B|C}$ and ${\mathcal{N}}_{AB|C}$
obeys the following inequality
\begin{equation}
{\mathcal{N}}^{2}_{AB|C} \geq
{\mathcal{N}}^{2}_{A|C}+{\mathcal{N}}_{B|C}^{2},
\label{eq:monogamy}
\end{equation}
where ${\mathcal{N}}_{A|C}$ and ${\mathcal{N}}_{B|C}$ are the
negativity between AC or BC, respectively. This is called the monogamy
relation of the negativity, which is an crucial property of the
quantum entanglement. The monogamy relation is proved for a
multi-qubit system \cite{Osborn2006}, and the similar relation holds
for the entanglement measure defined in the Gaussian system
\cite{Hiroshima2007}. However, there is no proof of the monogamy
relation of the negativity for the Gaussian system. If the monogamy
relation holds then the multipartite entanglement can be expressed by
the quantity
\begin{equation}
{\mathcal{N}}_{A|B|C}:={\mathcal{N}}^{2}_{AB|C}-{\mathcal{N}}^{2}_{A|C} - {\mathcal{N}}_{B|C}^{2}.
\end{equation}
This is the difference between the two side of Eq.~\eqref{eq:monogamy}
and can be interpreted as the residual entanglement. If this quantity
vanishes, the entanglement between AB and C can be decomposed to the
entanglement between A and C, and the entanglement between B and
C. Thus the entanglement between AB and C is written as sum of pure
bipartite entanglement between sub system. In such a case, there is no
multipartite entanglement.  

\begin{figure}[H]
  \centering
   \includegraphics[width=0.6\linewidth]{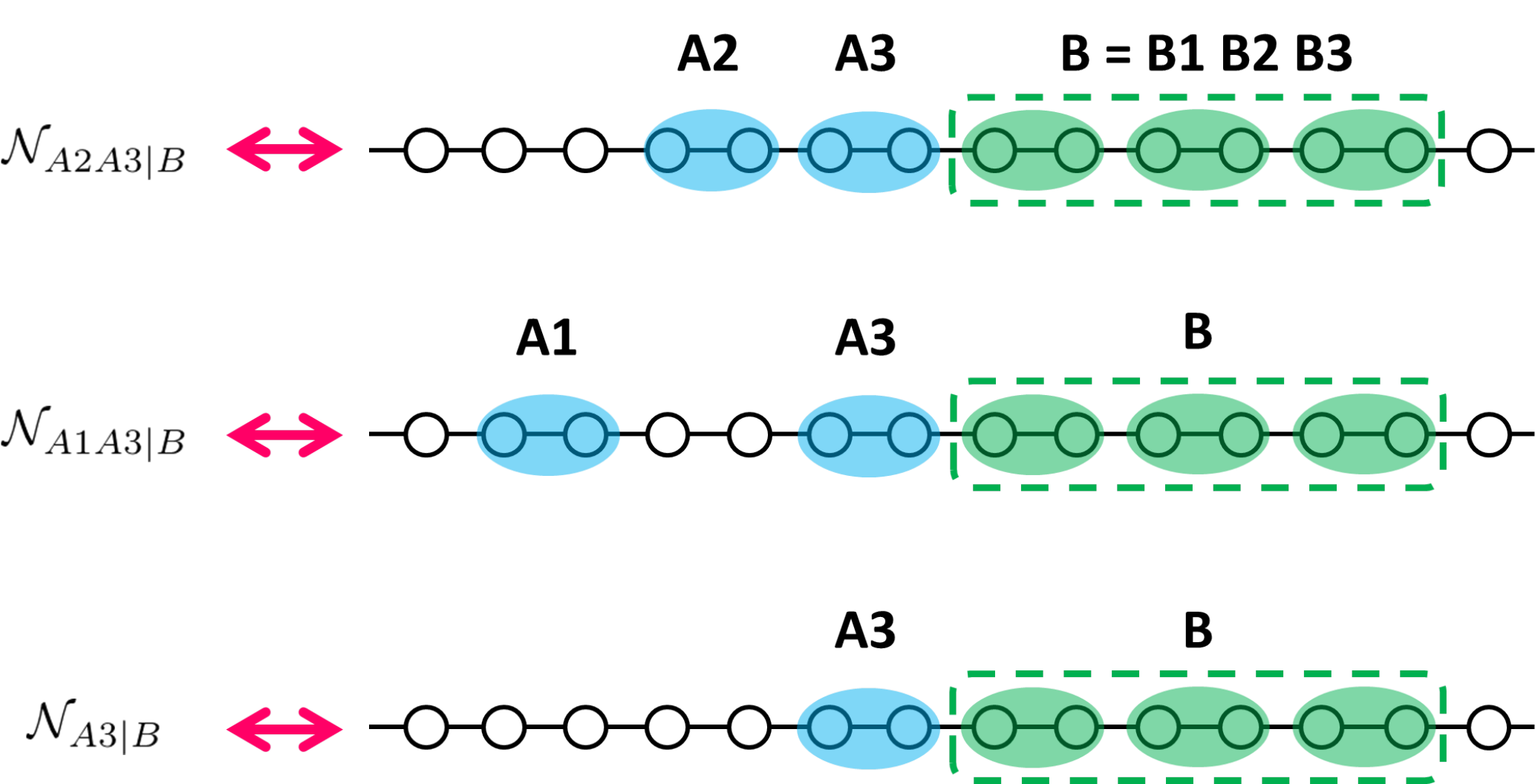}
   \caption{The correspondence between each negativity in the
     tripartite system and the two regions AB for $d=0$.}
   \label{fig3}
\end{figure}

\begin{figure}[H]
 \begin{minipage}{0.5\hsize}
   \centering
   \includegraphics[width=0.95\linewidth]{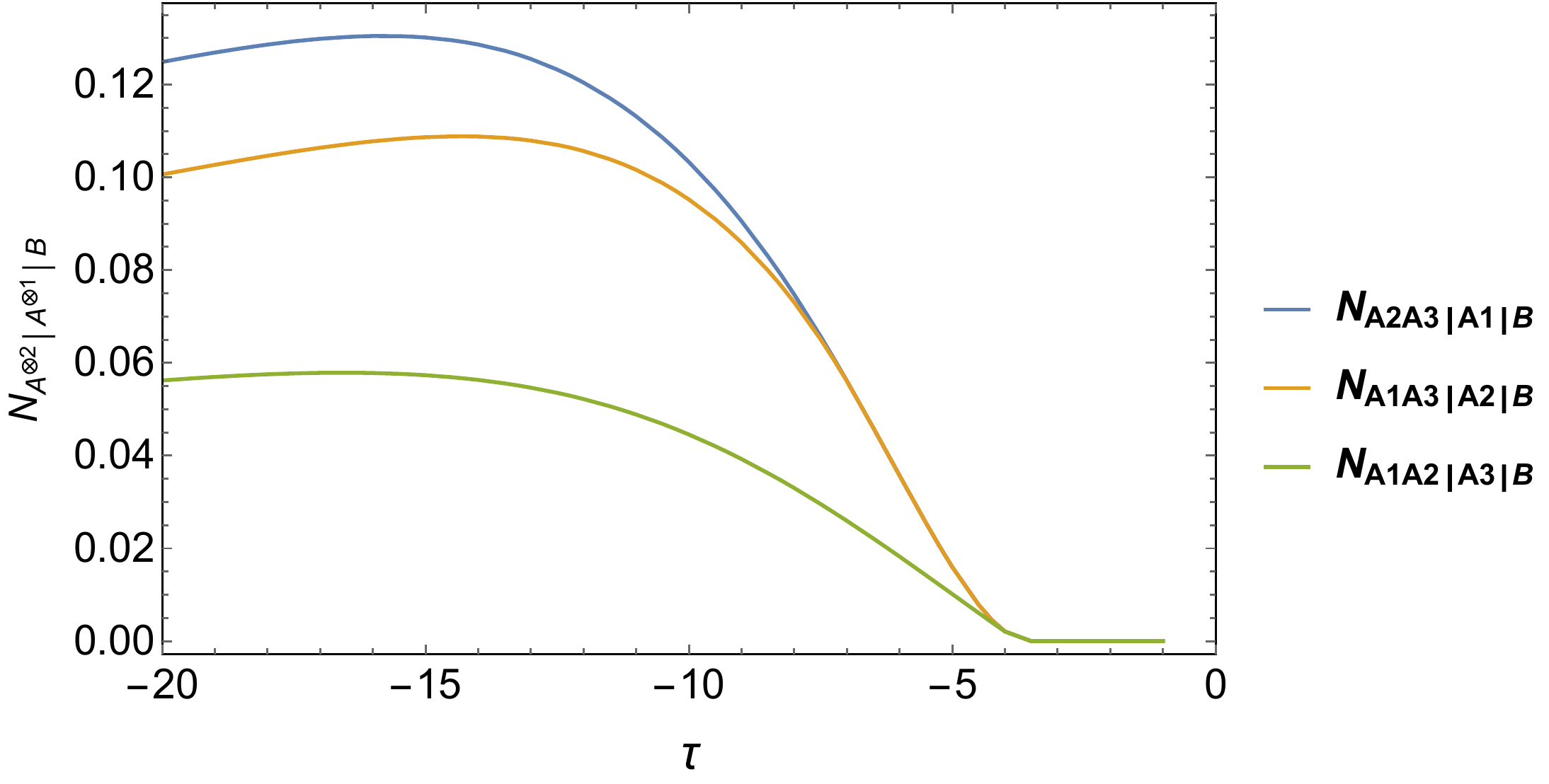}
 \end{minipage}
 \begin{minipage}{0.5\hsize}
   \centering
   \includegraphics[width=0.98\linewidth]{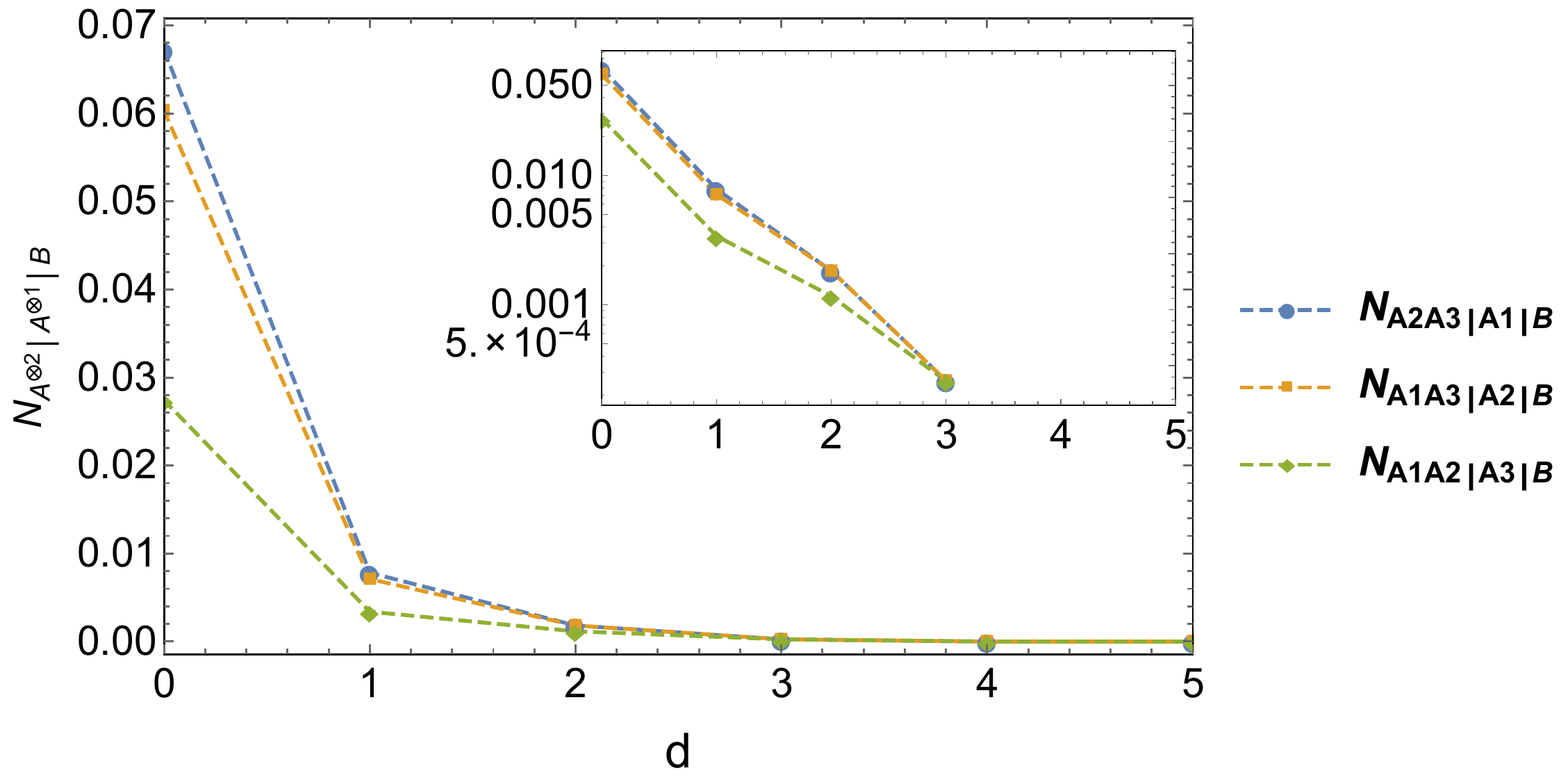}
 \end{minipage}
 \caption{Left panel: the time dependence of  ${\mathcal{N}}_{A^{\otimes 2}|A^{\otimes 1}|B}$ with $d=0$. 
   Right panel: the distance dependence of ${\mathcal{N}}_{A^{\otimes 2}|A^{\otimes 1}|B}$ at $\tau=-80$}
    \label{fig4}
\end{figure}

To understand the multipartite entanglement in the Bunch-Davies
vacuum, we consider the negativity with the number of each mode $n=3$
in the systems A and B (FIG.~\ref{fig3}).  The density operator
$\hat\rho_{AB}$ is a $3 \times 3$ mode Gaussian state and the system A is
composed of $3$ subsystems $\rm{A_{1}A_{2}A_{3}}$ (similarly, B is
$\rm{B_{1}B_{2}B_{3}}$). The monogamy relation is written as
\begin{equation}
{\mathcal{N}}^{2}_{A^{\otimes 3}|B} \geq {\mathcal{N}}_{A^{\otimes 2}|B}^{2}+{\mathcal{N}}^{2}_{A^{\otimes 1}|B}, \label{26}
\end{equation}
where $A^{\otimes 3}=A_{1}A_{2}A_{3}$ and
$(A^{\otimes 2},A^{\otimes
  1})=(A_{1}A_{2},A_{3}),~(A_{1}A_{3},A_{2}),~(A_{2}A_{3},A_{1})$.
For example, each negativity
${\mathcal{N}}_{A_{2}A_{3}|B},~{\mathcal{N}}_{A_{1}A_{3}|B}$ and
${\mathcal{N}}_{A_{3}|B}$ which appears on the right side for
(\ref{26}) corresponds to the entanglement between the two regions A
and B shown in FIG. \ref{fig3}.
\noindent
Let us check the monogamy relation of the negativity for the
Bunch-Davies vacuum in the case $n=3$. The left and right panel of FIG. \ref{fig4}
present the time dependence with $d=0$ and the distance dependence at
$\tau=-80$ of the quantity
${\mathcal{N}}_{A^{\otimes 2}|A^{\otimes 1}|B}$, respectively.
\noindent
From these results, we confirm that the monogamous relation of the
negativity ${\mathcal{N}}_{A^{\otimes 2}|A^{\otimes 1}|B}\ge 0$ holds for our model. Hence we can characterize the multipartite
entanglement in de Sitter space by the negativity.

\begin{figure}[H]
 \begin{minipage}{0.5\hsize}
   \centering
   \includegraphics[width=0.95\linewidth]{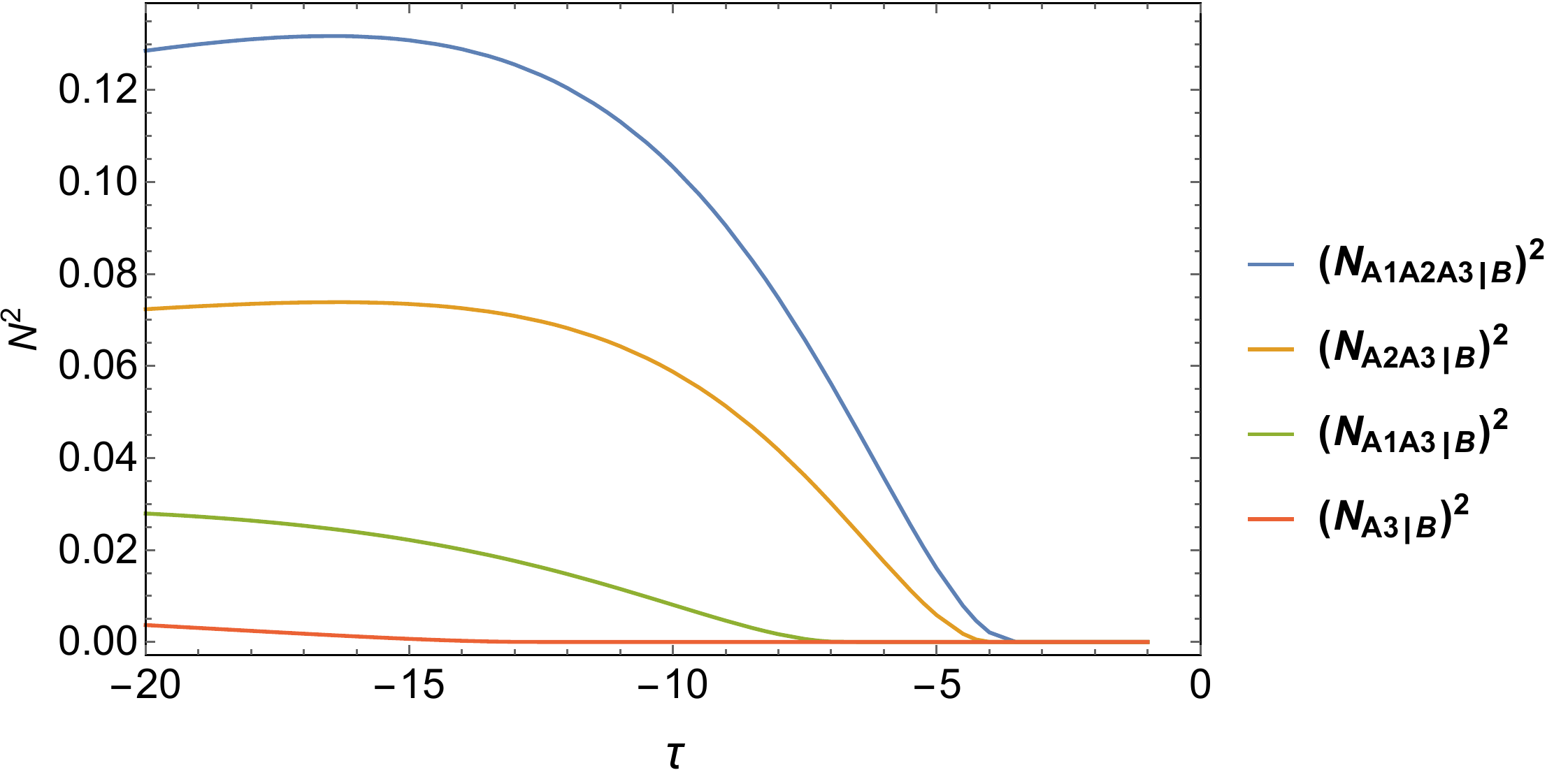}
 \end{minipage}
 \begin{minipage}{0.5\hsize}
   \centering
   \includegraphics[width=0.98\linewidth]{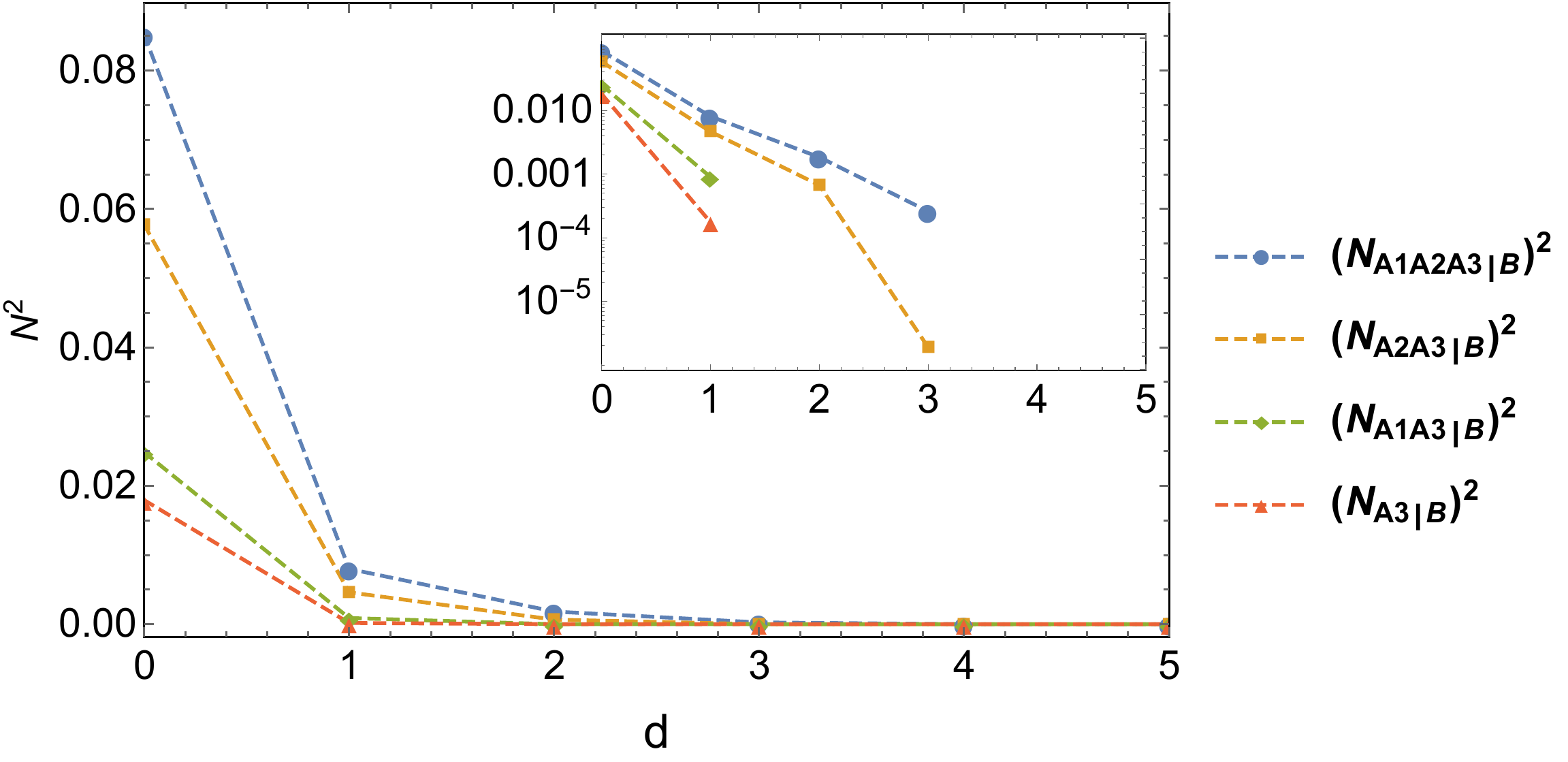}
 \end{minipage}
 \caption{Left panel: square of each negativity
   ${\mathcal{N}}_{A_{2}A_{3}|B},~{\mathcal{N}}_{A_{1}A_{3}|B}$ and
   ${\mathcal{N}}_{A_{3}|B}$ as a function of $\tau$ with $d=0$. 
   Right panel: square of these
   negativities as a function of $d$ at $\tau=-80$, and the upper
   inset is its log-plot.}
    \label{fig5}
\end{figure}

In the left panel of FIG. \ref{fig5}, the time dependence of
${\mathcal{N}}^{2}_{A_{1}A_{2}A_{3}|B},~{\mathcal{N}}^{2}_{A_{2}A_{3}|B},~{\mathcal{N}}^{2}_{A_{1}A_{3}|B}$
and ${\mathcal{N}}^{2}_{A_{3}|B}$ for $d=0$ is shown (the other
cases ${\mathcal{N}}_{A_{1}A_{2}|B},~{\mathcal{N}}_{A_{1}|B}$ and
${\mathcal{N}}_{A_{2}|B}$ are trivially zero).  We observe that the
negativities ${\mathcal{N}}^{2}_{A^{\otimes 1}|B}$ and
${\mathcal{N}}^{2}_{A^{\otimes 2}|B}$ decay faster than the negativity
${\mathcal{N}}^{2}_{A^{\otimes 3}|B}$ for the $3 \times 3$ mode
Gaussian system. This behavior guarantees the monogamy inequality
\eqref{eq:monogamy}.
\noindent 
The right panel of FIG. \ref{fig5} shows that the distance dependence
of
${\mathcal{N}}^{2}_{A^{\otimes 3}|B},~{\mathcal{N}}^{2}_{A^{\otimes
    2}|B}$
and ${\mathcal{N}}^{2}_{A^{\otimes 1}|B}$ for $\tau=-80$ (the other
cases ${\mathcal{N}}_{A_{1}A_{2}|B},~{\mathcal{N}}_{A_{1}|B}$ and
${\mathcal{N}}_{A_{2}|B}$ are trivially zero again). As in the case of
the left panel of FIG. \ref{fig5}, the negativities
${\mathcal{N}}^{2}_{A^{\otimes 1}|B}$ and
${\mathcal{N}}^{2}_{A^{\otimes 2}|B}$ decrease more than
${\mathcal{N}}^{2}_{A^{\otimes 3}|B}$ with the distance $d$ to keep
the monogamy relation \eqref{eq:monogamy}. The behaviors observed in 
FIG. \ref{fig5} also suggest that the multipartite entanglement remains in the 
super horizon scale when the number of modes $n$ becomes large.

\subsection{Continuous limit}
To investigate the entanglement for the super horizon scale, we
consider the continuous limit of our lattice model, where multipartite
entanglement plays an important role.  For realization of the
continuous limit of our lattice model, we use the canonical variables
with $n_{\rm{c}}=1$ (no coarse-graining), and choose each parameter as
$N=2\times 10^{4}$ and $\alpha=1-10^{-12}$, again. It is also assumed
that each region contains $l$ harmonic oscillators and their comoving
separation is $d$ (FIG. \ref{fig6}). In the appendix, we present the
convergence check and the small violation of the uncertain relation
due to numerical error to confirm that our numerical calculation
really corresponds to the continuous limit and is stable. We 
 compare the previous
works \cite{Marcovitch2009a,Nambu2008} with our numerical results. In
Ref~\cite{Marcovitch2009a}, the authors considered a massless scalar
field in the 1+1-dimensional Minkowski space and numerically showed
that the logarithmic negativity of a massless scalar field between two
spatially regions decays exponentially as the ratio $d/l$
increases. The property that the logarithmic negativity depends only
on the ratio $d/l$ is derived from the scale invariant for the
massless theory. In the following, we investigate the entanglement for
the super horizon scale and how it depends on the Hubble scale $H$.
\begin{figure}[H]
\centering
    \includegraphics[clip,width=0.7\linewidth]{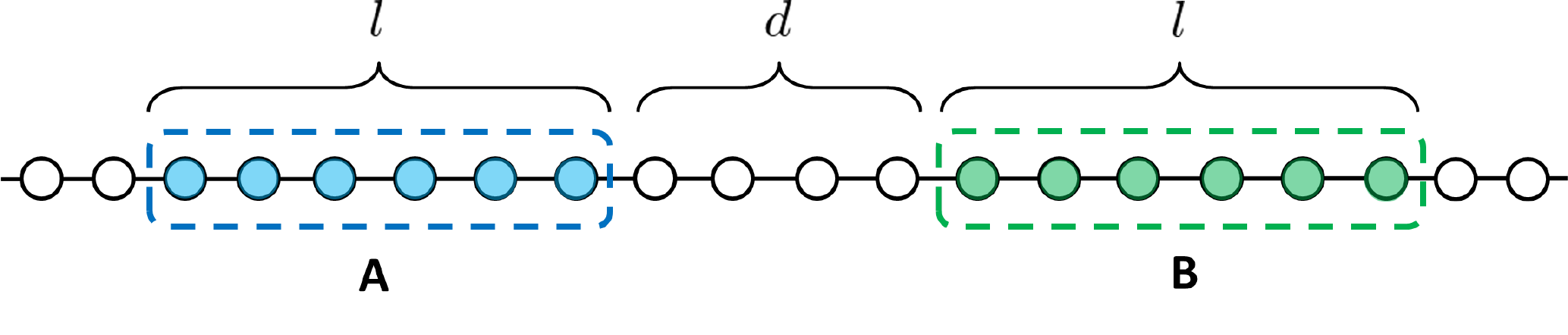}
    \caption{Two spatial symmetric regions A and B in the harmonic
      chain. $l$ is the comoving size of each region and $d$ is the comoving distance between the two regions. In this case, we consider the coarse-graining scale $n_{\rm{c}}=1$ to investigate the continuous limit of our model.}
    \label{fig6}
\end{figure}
In Ref.~\cite{Nambu2008}, the logarithmic negativity of the
coarse-grained field in de Sitter space vanishes when the two regions
are causally disconnected and FIG.~\ref{fig2} also shows the
  negativity with the coarse-grained field becomes zero for
  sufficiently large scales or late times. On the other hand,
the negativity obtained without coarse-graining
  (FIG.~\ref{fig7}) does not vanish even when the distance between
two regions is larger than the horizon scale ($d_p=1$ corresponds to
the Hubble horizon scale). This observation confirms that the
multipartite entanglement remains on the super horizon scale as
expected above. For a vacuum state in the quantum field theory, the
Reeh-Schrieder theorem characterizes the (multipartite) entanglement
of quantum field~\cite{Reeh1961}. Our numerical results suggest that
the Reeh-Schrieder theorem also holds for the Bunch-Davies vacuum in de
Sitter space.
\begin{figure}[H]
  \centering
   \includegraphics[width=0.6\linewidth]{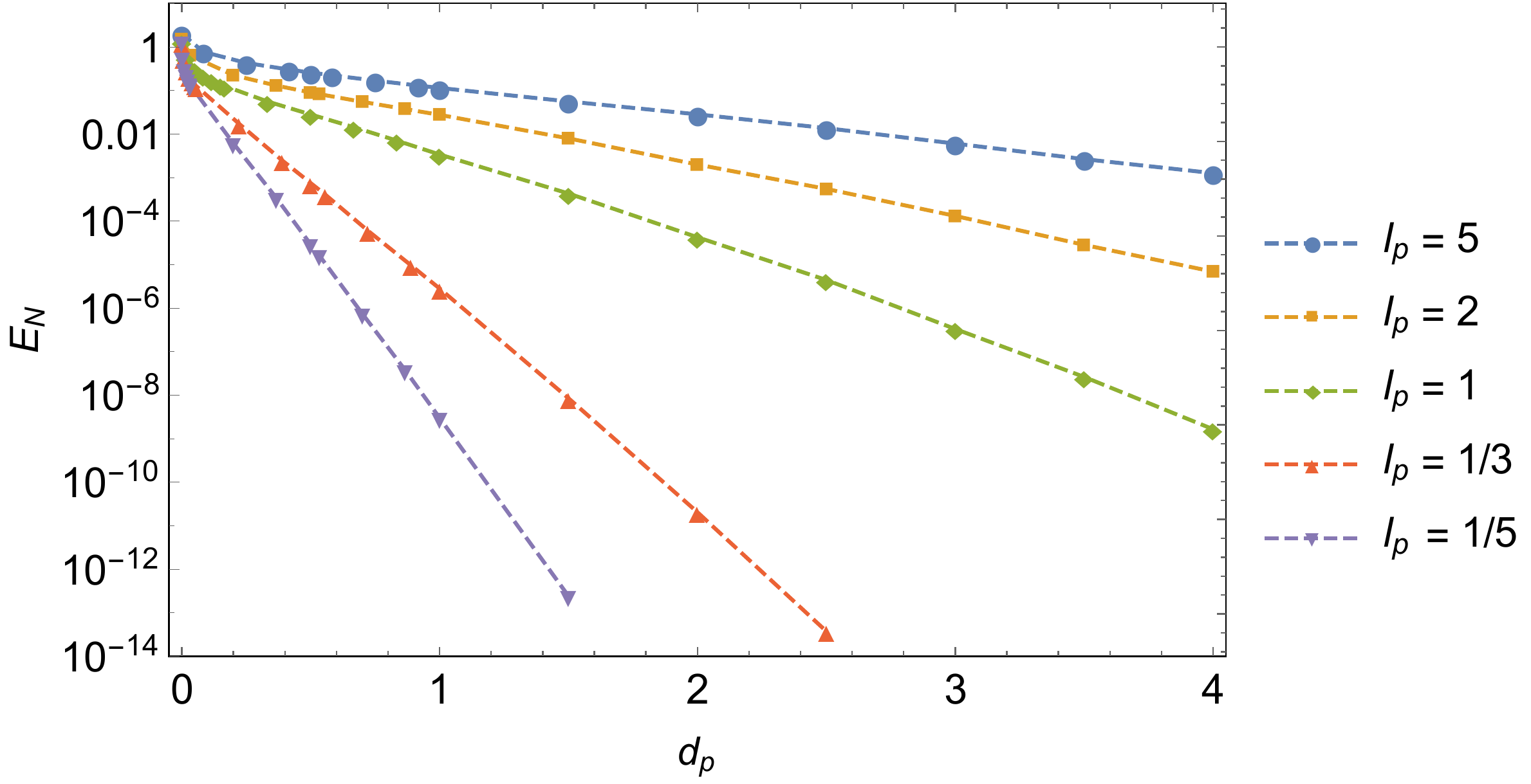}
   \caption{The behavior of the logarithmic negativity $E_{\mathcal{N}}$ as a
     function of $d_{p}$ with fixed $l_{p}$.}
   \label{fig7}
\end{figure}
As the Bunch-Davies vacuum approaches to the Minkowski vacuum in the
remote past, the behavior of the logarithmic negativity for $l_{p} <1$
and $d_{p} <1$ is expected to be same as that for the Minkowski
case. To focus on the entanglement peculiar to de Sitter space, we
consider the behavior of the logarithmic negativity for $l_{p} \geq1$
and $d_{p} \geq1$.  FIG.~\ref{fig8} shows the logarithmic negativity
$E_{\mathcal{N}}$ as a function of $d_{p}$ in this case.
\begin{figure}[H]
  \centering
   \includegraphics[clip,width=0.6\linewidth]{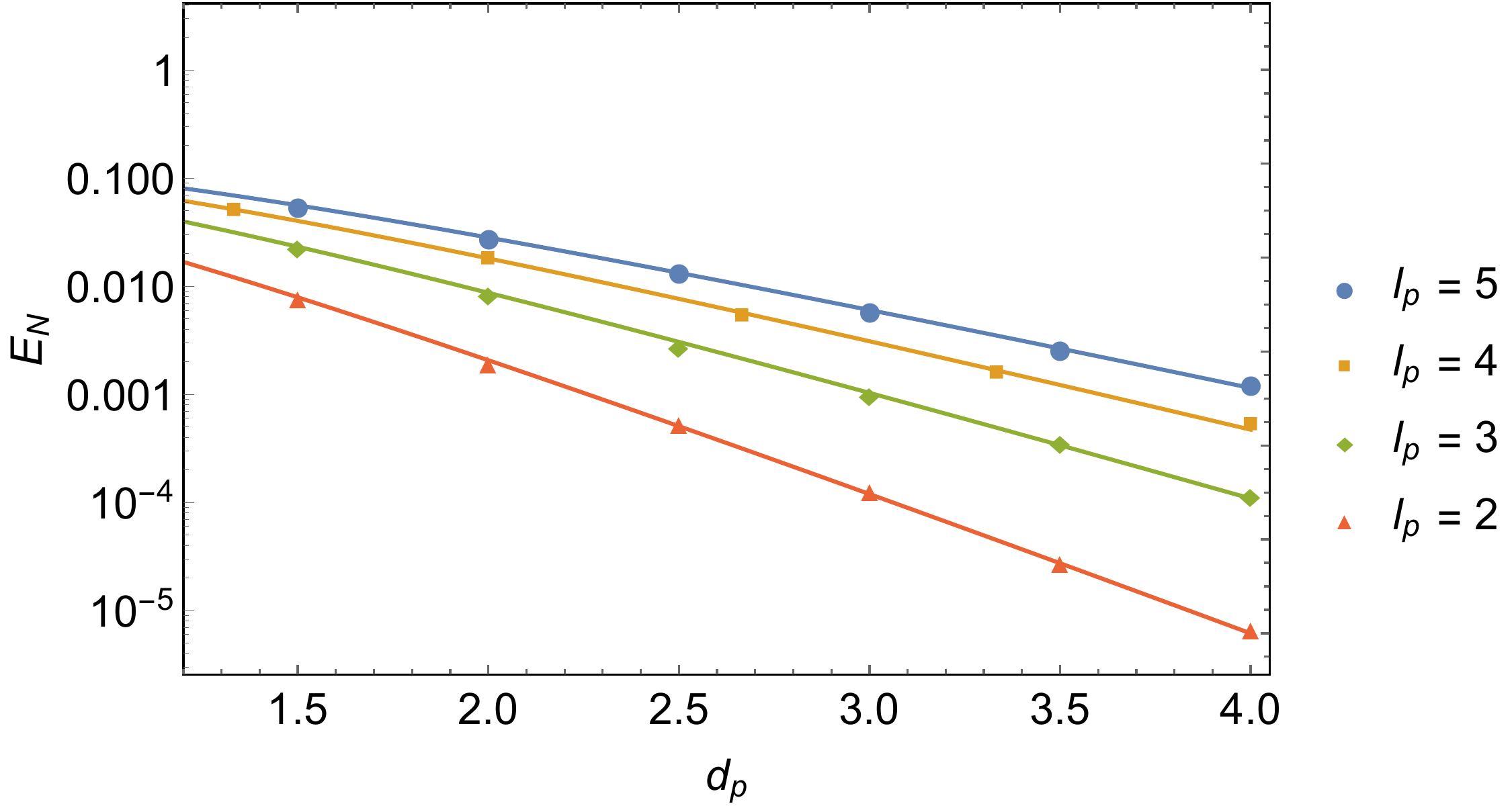}
   \caption{The behavior of the logarithmic negativity $E_{\mathcal{N}}$ as a
     function of $d_{p}\ge 1$ with fixed $l_{p}\ge 1$.}
   \label{fig8}
\end{figure}
\noindent
The logarithmic negativity for $l_{p} \geq1$ and $d_{p} \geq1$ behaves
as almost linear functions in log plot. We use the fitting
function of the exponential factor with the power-law correction to
compare with the logarithmic negativity in the Minkowski vacuum
\cite{Marcovitch2009a}. The solid lines in FIG.~\ref{fig8} represent
the fitting result
\begin{equation}
E_{\mathcal{N}}^{\text{(fit)}} \approx d_{p}\,e^{-k\,d_{p}},
\end{equation}
where $k$ is a real parameter whose values depend on the ratio
$l_{p}$. FIG.~\ref{fig9} shows $k$ as a function of $l_{p}$ and the
solid line in the figure represents a function $k=a_{1}+a_{2}\,l_{p}^{-1}$
where $a_{1}$ and $a_{2}$ are  $O(1)$ constants given by
$a_{1} \sim1.08$ and $a_{2}\sim 4.35$ . 
\begin{figure}[H]
  \centering
   \includegraphics[clip,width=0.5\linewidth]{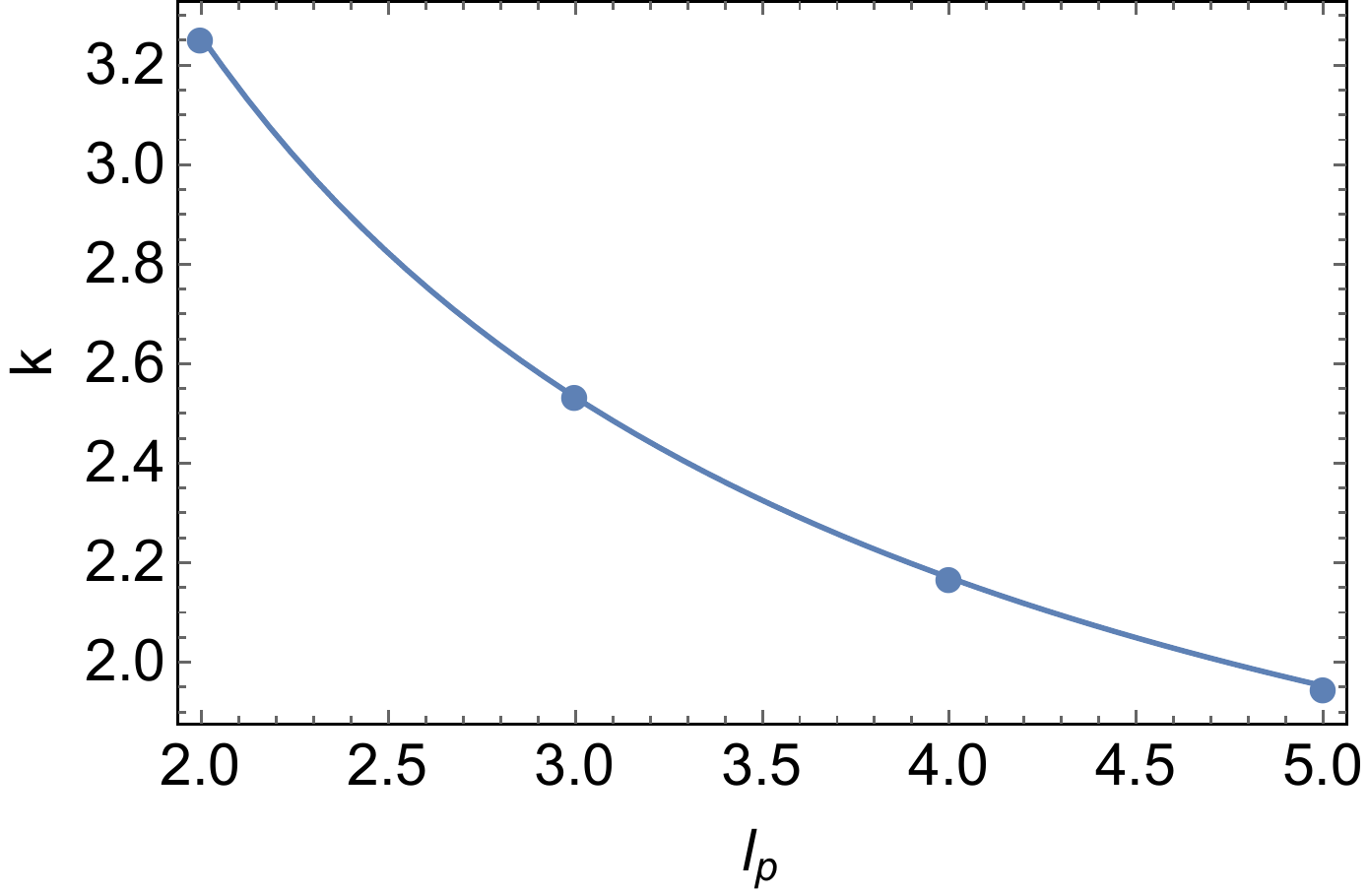}
   \caption{The coefficient $k$ of $d_{p}$ as a function of $l_{p}$.}
   \label{fig9}
\end{figure}
\noindent
Combining these, we obtain the following fitting formula of the
logarithmic negativity for $d_{p} \geq 1$ and $l_{p} \geq 1$:
$ E_{\mathcal{N}}^{\text{(fit)}} \approx d_{p}\,e^{-a_{1}d_{p}-a_{2}
  (d_{p}/l_{p})}$.
By restoring the dimension of the variables, this is rewritten as
\begin{equation}
E_{\mathcal{N}}^{\text{(fit)}} \approx \frac{D_p}{H^{-1}} \exp
\left[-a_{1}\frac{D_p}{H^{-1}}-a_{2}\frac{D_p}{L_p} \right], \label{31}
\end{equation}
where $H^{-1}$ is the Hubble length and the proper size of the region
$L_p$ and distance $D_p$ are given by
$L_p=a(\eta)\,l\, \Delta x,~D_p=a(\eta)\,d\, \Delta x$. In the
  formula \eqref{31} we observe that the logarithmic negativity decays
  exponentially with respect to $H$. This means that the quantum
  entanglement is degraded by the thermal noise with the Hawking
  temperature $H$. Thus we consider the value of the numerical factor
  $a_{1}$ is related to the thermal noise which is independent of
  details of the theory. On the other hand, the interpretation of the
  factor $a_{2}$ is not so clear because a value of this coefficient
  of $D_p/L_p$ depends on the theory. To make clear its physical
  interpretation, we will need further analysis of quantum
  entanglement using other theories of the scalar field in de Sitter
  space. For the super horizon regions $L_p \gg H^{-1}$, the
logarithmic negativity \eqref{31} becomes independent of the size
$L_p$ of the two symmetric spatial regions and its value is determined
only by the ratio $D_p/H^{-1}$. We expect that this property can be
understood as follows: the physical wavelength of quantum fluctuation
in the considering regions (FIG.~\ref{fig6}) is initially smaller than
the Hubble horizon. As the universe expands, the wavelength exceeds
the Hubble horizon. After the horizon exit, the scale of fluctuations
in each region is determined only by the Hubble horizon scale. Hence,
the amount of entanglement depends only on $D_{p}/H^{-1}$.

The above property of the logarithmic negativity (\ref{31}) 
is expected to be true for the 1+3-dimensional de Sitter space. This
is because the feature of the entanglement is determined by the mode
function and we use the same mode function as the 1+3-dimensional
model. As the quantum fluctuation of super horizon mode is scale
independent, the quantum entanglement is determined only by the Hubble scale and the separation between the considering regions. 
\section{Summary and conclusion}

We investigated the quantum entanglement between two symmetric spatial
regions with the Bunch-Davies vacuum for the 1+1-dimensional effective
harmonic chain model. We introduced the coarse-grained variables and
examined the multipartite entanglement in de Sitter spacetime by the
monogamy relation of the negativity. In the previous work
\cite{Nambu2008}, it has been shown that the bipartite entanglement
disappears on the super horizon scale. In contrast, in this paper, it
was found that the multipartite entanglement of the super horizon
scale remains. This indicates that the multipartite entanglement plays
an important role to characterize the quantum nature of the the super
horizon scale fluctuations.  We also considered the continuum limit of
our model and calculated the logarithmic negativity for the original
canonical variables (without coarse-graining). We confirmed that the
logarithmic negativity remains non-zero even if the distance between
the two regions becomes larger than the Hubble length. That is, there
exists the quantum entanglement between two causally disconnected
regions and the existence of the entanglement means that the
Reeh-Schrieder theorem holds in de Sitter space.

Finally, we comment on the relation between our lattice model and
1+3-dimensional theory. We considered the 1+1-dimensional effective
lattice model of the free massless scalar field in the de Sitter
space. As the behavior of the logarithmic negativity depends on the
spatial dimension, the numerical simulation in our model is not
equivalent to the universe with three spatial dimensions. At the
  end of Sec.~II, we observed that the two-point correlation in our
model is larger than it in a 1+3-dimensional de Sitter
space. Hence for the scalar field without coarse-graining
  (continuous limit), if the entanglement disappears for some size and
  separation of each region in 1+1-dimensional model, we expect that
  the entanglement in the corresponding 1+3-dimensional model also
  vanishes. However, according to our numerical calculation, the
  entanglement in 1+1-dimensional model in the continuous limit exists
  in any scale.  Thus our lattice model provides the necessary
condition to judge the existence of quantum entanglement in the
1+3-dimensional de Sitter space. The main features of the logarithmic
negativity found in our analysis is characterized by properties of the
mode function of the scalar field in de Sitter space. The equation in
our model is the same as that in the 1+3-dimensional de Sitter
spacetime, and if we evaluate quantum entanglement between spatial
regions in the 1+3-dimensional de Sitter spacetime, we expect that we
will obtain the similar feature or property of entanglement obtained
this paper.

\begin{acknowledgments}
This work was supported in part by the JSPS KAKENHI Grant Number 16H01094.
\end{acknowledgments}
\newpage
\begin{appendix}

\begin{figure}[H]
  \centering
  \includegraphics[clip,width=0.7\linewidth]{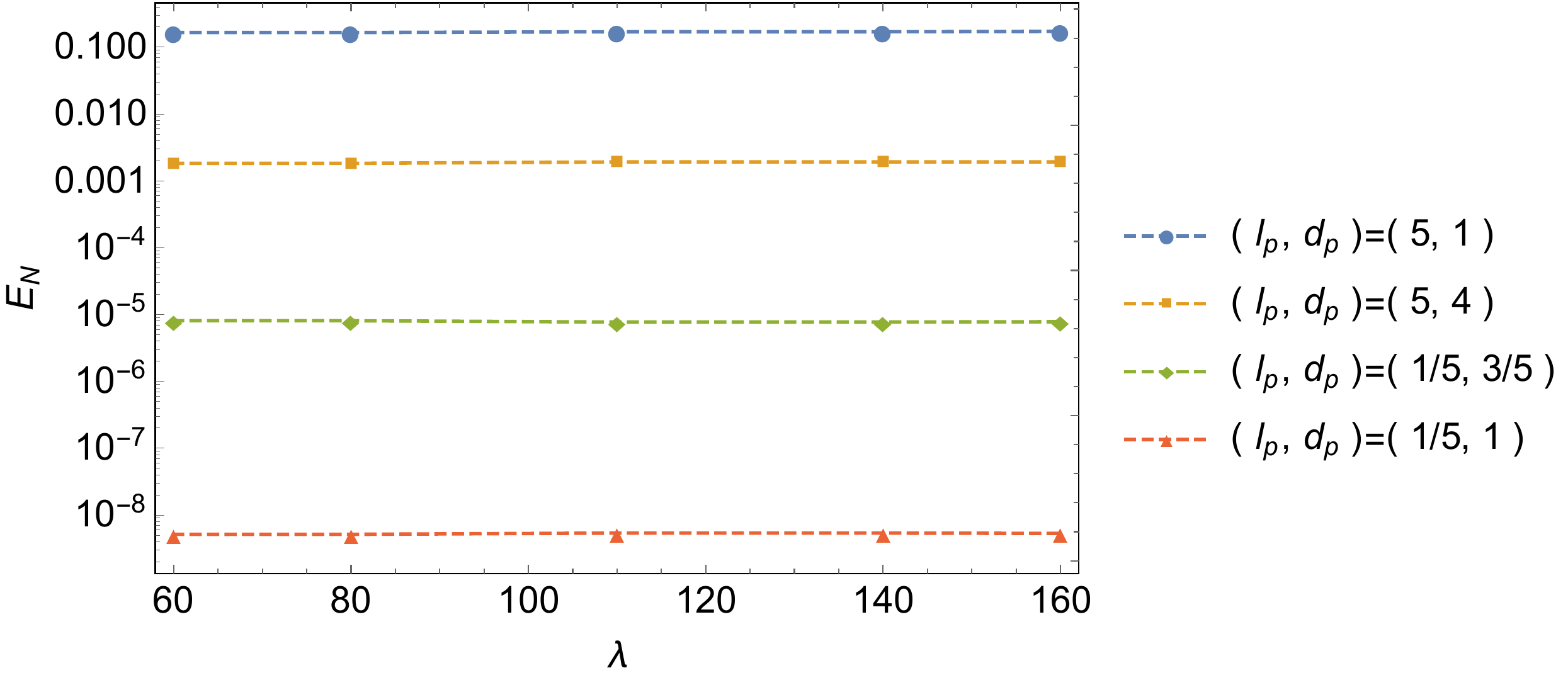}
  \caption{The convergence check of $E_{\mathcal{N}}$ for
    $(l_{p},d_{p})=(5,1),(5,4),(1/5,3/5),(1/5,1)$. The
      continuum limit corresponds to $\lambda\rightarrow\infty$.}
  \label{fig10}
\end{figure}

{\flushleft{\section{Convergence check and violation of uncertainty relation}}}
To confirm that our numerical calculation corresponds to the continuum
limit of the lattice model, we check the convergence of the
logarithmic negativity. By introducing a scaling parameter $\lambda$,
we write other parameters contained in the model as
\begin{equation}
 N=200\times\lambda,\quad
 \alpha=1-\lambda^{-2} \times 10^{-8},
 \quad l=\lambda.
\end{equation}
We regard $E_{\mathcal{N}}$ as a function of $\lambda$ for fixed $l_{p}$ and
$d_{p}$. $\lambda\rightarrow\infty$ corresponds to the continuum limit
of the model. FIG.~\ref{fig10} shows the result of convergence check.
\noindent
The logarithmic negativity is independent of the parameter
$\lambda$. In the limit $\lambda\gg 1$, length scales (for example,
the physical size of the considering region or the distance) is much
larger than the UV cutoff $\Delta x$. Hence, this limit
corresponds to the continuum limit and our numerical calculation well
approaches the continuum limit.

\begin{figure}[H]
  \centering
    \includegraphics[clip,width=0.65\linewidth]{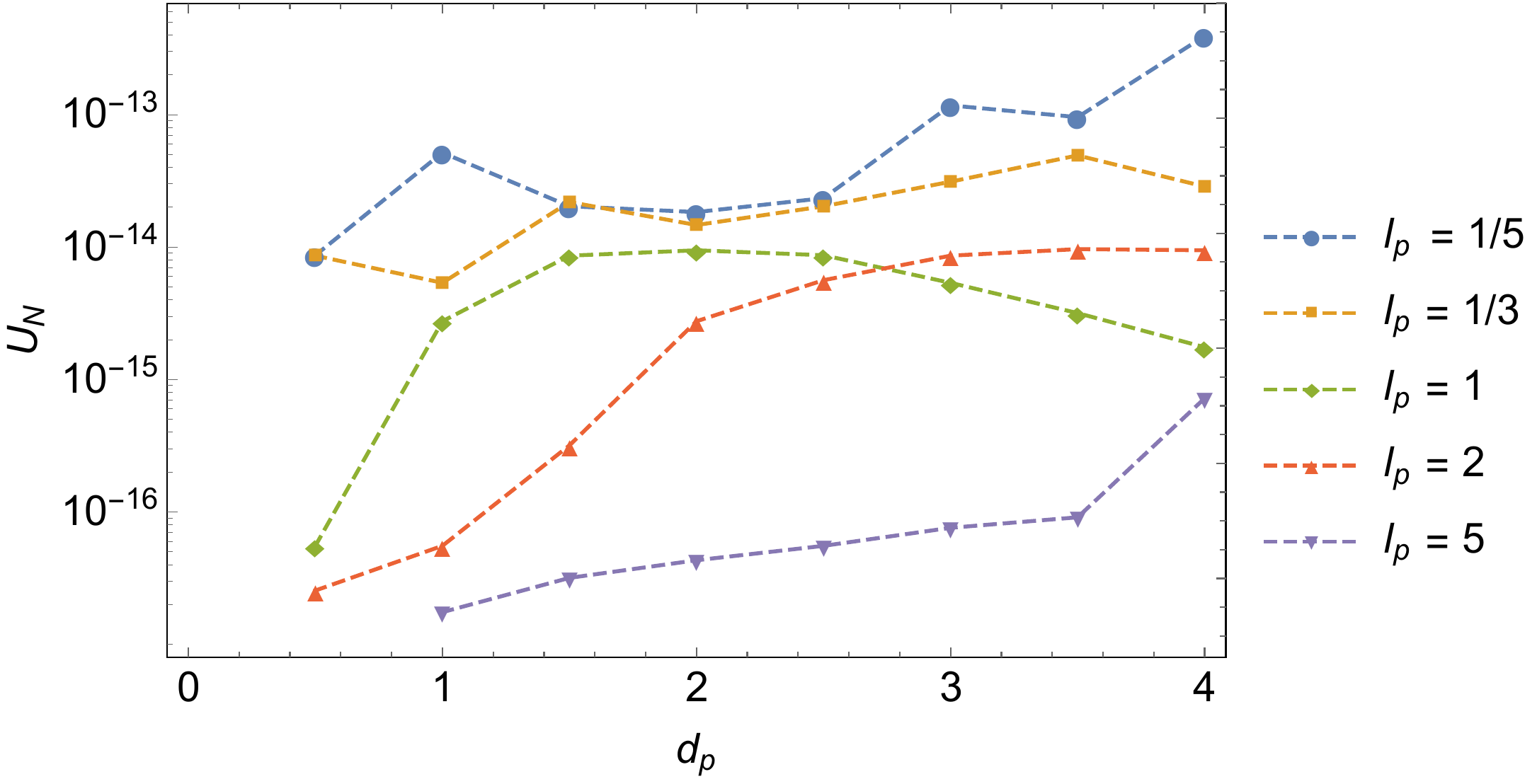}
  \caption{The violation of the Heisenberg uncertainty relation for
    the fixed ratio $l_{p}=5,2,1,1/3,1/5$ in our numerical
    calculation. The uncertainty relation holds if $U_N=0$ is satisfied.}
  \label{fig11}
\end{figure}

Furthermore, we evaluate  violation of the Heisenberg uncertainty
relation for our numerical calculation by checking a quantity defined
by
\begin{equation}
U_{N}=-\sum_{j=1}^{N} \log_{2} \left[\min{\left(2{\nu}_{j},1 \right) }\right],
\end{equation}
where ${\nu}_{j}$ are eigenvalues of $i\,\Omega V_{AB}$. If $U_{N}=0$
then we get relations $\nu_{j} \geq 1/2$, which are equivalent
to the uncertainty relation. FIG.~\ref{fig11} shows $U_{N}$ as a
function of $d_{p}$ for fixed $l_{p}$.
\noindent
The uncertainty relation is expressed in terms of the two point
functions $\langle \hat q\hat q\rangle,\langle \hat p\hat p\rangle$
and $\langle \hat q\hat p+\hat p\hat q\rangle$.  For the massless
theory, the $\hat q\hat q$-correlation has the IR divergence. On the
other hand, there is no the IR divergence in
$\hat p\hat p,\hat q\hat p+\hat p\hat q$-correlations. In our
numerical calculation, owing to behavior of the mode function in de
Sitter space, there appears the large difference of the magnitude
between the $\hat q\hat q$- and
$\hat p\hat p,\hat q\hat p+\hat p\hat q$-correlations. The violation
of the uncertainty relation due to numerical error tends to become
larger as $d_{p}$ increases.  According to FIG.~\ref{fig11}, this
violation of the Heisenberg uncertainty relation is kept small enough
to guarantee accuracy of our numerical calculation.
\end{appendix}


\end{document}